%% file: cog.tex
\documentclass[prl,reprint]{revtex4-1}
\usepackage{graphicx}
\usepackage{epsfig}
\usepackage{fancyhdr}
\usepackage{mathbbol}
\usepackage{pstricks}
\usepackage{color}
\usepackage{bbm}
\usepackage{multirow}
\usepackage{amsmath}
\usepackage{array} 
\usepackage{tabularx}
\usepackage{tabulary}
\usepackage{dcolumn} 
\usepackage{nccmath} 
\include{definitions}

\begin{document}

\title{How the small hyperfine splitting of P-wave mesons evades large loop corrections}

%

\author{T. J. Burns }
\email{{\tt Timothy.Burns@roma1.infn.it}\href{mailto:Timothy.Burns@roma1.infn.it}}
\affiliation{INFN Roma, Piazzale A. Moro 2, Roma, I-00185, Italy}
\begin{abstract}
The recent discoveries of the bottomonia states $h_b(1\uP)$ and $h_b(2\uP)$ confirm the quark model prediction, already verified in the charmonia sector, that the hyperfine splitting of P-wave mesons is very small. The striking agreement is somewhat surprising because the nonrelativistic result, for which the splitting is zero, may be modified due to large mass shifts from coupling to open flavour meson pairs. This paper is based on the observation that in most models hyperfine splitting remains small despite what are in many cases large mass shifts. This effect is shown to be a generic feature of models in which the coupling is driven by the creation of a light quark pair with spin-one. 
\pacs{12.39.Jh,12.39.Pn,12.40.Yx,14.40.Pq}
\end{abstract}


\maketitle

\thispagestyle{fancy}

Quark potential models share a common spin-dependent structure, which in perturbation theory yields an expression for the mass $M_{SLJ}$ of a state of spin $S$, orbital angular momentum $L$ and total angular momentum $J$,
\bea
M_{SLJ}	&=&M+\Delta_{SLJ},\label{mass}\\
\Delta_{SLJ}&=&
\Delta_{s}\evhh_S
+\Delta_{t}\evt_{SLJ}
+\Delta_{o}\evso_{SLJ},
\nonumber
\eea
in terms of common expectations values $M$, $\Delta_{s}$, $\Delta_{t}$ and $\Delta_{o}$ of the spin-independent, spin-spin, tensor and spin-orbit terms. For states with $L\ne 0$, certain linear combinations of masses are independent of three of the four expectation values, and the most interesting of these is the hyperfine splitting. For the P-wave family it is expressed:
\be
\frac{1}{9}\left(M_{\an 3P0}+3M_{\an 3P1}+5M_{\an 3P2}\right)-M_{\an 1\uP1}=\Delta_{s}.\label{zerohyperfine}
\ee
In the nonrelativistic limit $\Delta_{s}=0$, and the experimental charmonia masses are in excellent agreement with this prediction \cite{Dobbs:2008ec}:
\bes
\overline M_{\chi_c(1\uP)}-M_{h_c(1\uP)}=+0.02\pm 0.19\pm 0.13 \MeV.
\ees
Recently the BaBar collaboration discovered the $h_b(1\uP)$ in the decay $\Upsilon(3S)\to h_b\pi^+\pi^-$ \cite{:2011zp}, while Belle observed the same state, and discovered its radial excitation $h_b(2\uP)$, in the process $\epem\to h_b\pi^+\pi^-$  \cite{Adachi:2011ji}. The  corresponding splittings are also very small:
\beas
\overline M_{\chi_b(1\uP)}-M_{h_b(1\uP)} &=& +2\pm 4\pm 1 \MeV \textrm{(BaBar)},\\
\overline M_{\chi_b(1\uP)}-M_{h_b(1\uP)} &=& +1.62\pm 1.52 \MeV \textrm{(Belle)},\\
\overline M_{\chi_b(2\uP)}-M_{h_b(2\uP)} &=& +0.48^{+1.57}_{-1.22} \MeV \textrm{(Belle)}.
\eeas
Lattice QCD calculations likewise exhibit very small hyperfine splittings \cite{Gray:2005ur,Burch:2009az,Meinel:2010pv}. 

That the hyperfine splitting is so small is a triumph of the nonrelativistic quark model, but it is also something of a surprise. Relativistic effects, as well as admixtures of different orbital or radial components in any of the wavefunctions, will cause deviations from the nonrelativistic result. Moreover, it may be anticipated that mass shifts due to coupling to open flavour meson pairs will lead to further discrepancy. The aim of this paper is to show that the latter effect is small.

The effect of ``unquenching'' is to shift the masses of physical states downwards with respect to their bare ``quenched'' masses. Owing to their different spin and total angular momentum quantum numbers, states of given multiplet (such as $\chi_0$, $\chi_1$, $\chi_2$ and $h$) have different couplings, and the resulting mass shifts cause deviations from the quenched mass formula \rf{mass}. 

Barnes and Swanson  \cite{Barnes:2007xu} derive a loop theorem in the limit in which mesons sharing the same flavour, orbital and radial quantum numbers have equal mass. Remarkably, all such states are shifted by the same amount, so that the effect of unquenching can be absorbed into a redefinition of model parameters. In this  limit the equal mass shifts cancel exactly in equation \rf{zerohyperfine} and the nonrelativistic result of zero hyperfine splitting is maintained. 
\setlength{\tabcolsep}{-2pt}
\begin{table}
\begin{ruledtabular}
\begin{tabular}{l D{.}{.}{-1}		D{.}{.}{-1} 		D{.}{.}{-1}		D{.}{.}{-1}		D{.}{.}{-1}}

		&\multicolumn{1}{c}{$\Delta M_{\an 3P0}$}	
					&\multicolumn{1}{c}{$\Delta M_{\an 3P1}$}
								&\multicolumn{1}{c}{$\Delta M_{\an 3P2}$}
											&\multicolumn{1}{c}{$\Delta M_{\an 1\uP1}$}
														&\multicolumn{1}{c}{$\textrm{Ind.}$}\\
\hline
BS(1\uP,$\cc$)	&459			&496			&521			&504			&-1.8			\\
K (1\uP,$\cc$)	&198			&215			&228			&219			&-1.3			\\
LMC (1\uP,$\cc$)&35			&38			&63			&52			&-2.9			\\	
YLCD (1\uP,$\cc$)&131			&152			&175			&162			&-0.4			\\
OT (1\uP,$\cc$)	&173			&180			&185			&182			&0.0			\\
OT (1\uP,$\bb$)	&43			&44			&45			&44			&-0.4			\\
OT (2\uP,$\bb$)	&55			&56			&58			&57			&0.0			\\
LD (1\uP,$\bb$) 	&80.777			&84.823			&87.388			&85.785			&-0.013			\\
LD (2\uP, $\bb$)	&73.578			&77.608			&80.146			&78.522			&-0.048
\end{tabular}
\end{ruledtabular}
\caption{The magnitudes of the mass shifts computed in various models. The final column ``Ind.'' shows the induced hyperfine splitting due to loop effects.}
\label{shiftstable}
\end{table}

The same authors consider the more realistic case of mass shifts due to loops of $\uD$, $\uD^*$, $\uD_s$ and $\uD_s^*$ with physical masses, using also physical masses for the various charmonia. In principle there will be further corrections due to the different masses in loops with excited mesons, such as $\uD_0$, $\uD_1$, $\uD_1'$ and $\uD_2$, but it is reasonable to expect that these are smaller since they are suppressed by an energy denominator. The  mass shifts $\Delta M_{SLJ}$ calculated in this way are shown in the first row (BS) of Table \ref{shiftstable}. 

Although the relative shift between any two of the states is small ($\approx 50$ MeV) compared to the overall mass shifts ($\approx 500$ MeV), it is still very large compared to the scale of the experimental hyperfine splittings. It is thus striking to note that the correction to equation \rf{zerohyperfine} due to unquenching,
\be
-\frac{1}{9}\left(\Delta M_{\an 3P0}+3\Delta M_{\an 3P1}+5\Delta M_{\an 3P2}\right)+\Delta M_{\an 1\uP1},
\label{inducedhyperfine}
\ee
is just $-1.8$ MeV. 

Could this be accidental? For comparison Table \ref{shiftstable} presents the mass shifts and induced hyperfine splittings of charmonia and bottomonia 1P and 2P states in various other unquenched quark models: those of Kalashnikova (K) \cite{Kalashnikova:2005ui}, Li, Meng and Chao (LMC) \cite{Li:2009ad}, Yang, Li, Chen and Deng (YLCD) \cite{Yang:2010am}, Ono and T\"ornqvist (OT) \cite{Ono:1983rd}, and Liu and Ding (LD) \cite{Liu:2011yp}. Although the models differ markedly in the magnitude of the shifts, they share the common feature that the induced hyperfine splitting is in each case significantly smaller than relative mass shifts among the states, which are themselves significantly smaller than the overall mass shifts. The same, however, cannot be said of states which lie above open flavour threshold, and all of the calculations that follow apply only to subthreshold states. 

The effect is particularly noteworthy given the different assumptions underlying the various models. While the shifts in the approach of BS are derived in second order perturbation theory, those of K, LMC, YLCD, OT and LD are obtained by solving the coupled-channel equations. While BS, K and LD use harmonic oscillator wavefunctions  (either of universal size or with different sizes for charmed and charmed-strange mesons), LMC, YLCD and OT use wavefunctions obtained by solving a coulomb plus linear potential model. In BS, YLCD and OT, the pair creation strength is flavour-independent, while for K, LMC, and LD the creation of strange quarks is suppressed with respect to that of light quarks. 

One feature common to all models is that the coupling is driven by the creation of light quark pair in spin triplet. The quark spin and spatial degrees of freedom factorise so that the amplitude for the coupling can be expressed as a linear combination of spatial matrix elements, which are the overlaps of the meson spatial degrees of freedom, weighted by angular momentum recoupling factors. For  a state with $S$, $L$ and $J$ quantum numbers coupling to a pair of S-wave mesons with spins $s_1$, $s_2$, there is a single spatial matrix element $A_l$ for each partial wave $l$ \cite{Burns:2007hk}. The corresponding recoupling coefficients $C_{SLJ}^{s_1s_2l}$ can be derived from the general expression of ref. \cite{Burns:2006rz} and for the case of P-wave mesons are shown in Table \ref{metable}. 

The magnitude of the mass shift due to a given channel is
\be
\Delta M_{SLJ}^{s_1s_2l}	=C_{SLJ}^{s_1s_2l}\int dp \frac{p^2 |A_l(p)|^2}{\epsilon_{SLJ}^{s_1s_2}+{p^2}/{2\mu_{s_1s_2}}},\label{massshift}
\ee
while the probability (in perturbation theory) that the physical state is in the corresponding two-meson channel is
\be
P_{SLJ}^{s_1s_2l}		=C_{SLJ}^{s_1s_2l}\int dp \frac{p^2 |A_l(p)|^2}{\left(\epsilon_{SLJ}^{s_1s_2}+{p^2}/{2\mu_{s_1s_2}}\right)^2}.\label{probability}
\ee
The energy denominators are written in nonrelativistic form and are functions of the reduced mass and binding energy,
\bea
\mu_{s_1s_2}&=&\frac{m_{s_1}m_{s_2}}{m_{s_1}+m_{s_2}},\\
\epsilon_{SLJ}^{s_1s_2}&=&m_{s_1}+m_{s_2}-M_{SLJ},
\eea
for loop mesons with mass $m_{s_1}$ and $m_{s_2}$. If one uses the physical mass for $M_{SLJ}$ then \rf{massshift} is a coupled-channel equation; if instead one uses the bare mass for $M_{SLJ}$ then  \rf{massshift} is the second order perturbation to the energy shift.

The total mass shift and continuum probability are the sums over the corresponding quantities for the different spin channels and partial waves:
\be
\Delta M_{SLJ}=\sum_{s_1s_2l}\Delta M_{SLJ}^{s_1s_2l},\quad
P_{SLJ}=\sum_{s_1s_2l}P_{SLJ}^{s_1s_2l}.
\label{spinsums}
\ee

\begin{table}
\begin{ruledtabular}
\begin{tabular}{>{$}r<{$} >{$}c<{$}>{$}c<{$}>{$}c<{$}>{$}c<{$}}
		&{\an 3P0}	& {\an 3P1}	&{\an 3P2}	&{\an 1\uP1}	\\
\hline
C_{SLJ}^{00\uS}	&3/4		&0		&0		&0		\\
2C_{SLJ}^{10\uS}&0		&1		&0		&1/2		\\
C_{SLJ}^{11\uS}	&1/4		&0		&1		&1/2		\\
\hline
\sum_{s_1s_2}\widetilde{C}_{SLJ}^{s_1s_2\uS}
		&-1		&-1/2		&1/2		&0		\\
\hline
C_{SLJ}^{00\uD}	&0		&0		&3/20		&0		\\
2C_{SLJ}^{10\uD}&0		&1/4		&9/20		&1/2		\\
C_{SLJ}^{11\uD}	&1		&3/4		&2/5		&1/2		\\
\hline
\sum_{s_1s_2}\widetilde{C}_{SLJ}^{s_1s_2\uD}
		&1/2		&1/4		&-1/4		&0		\\
\end{tabular}
\end{ruledtabular}
\caption{Angular momentum recoupling factors.}
\label{metable}
\end{table}


It is reasonable to assume that mesons which share the same orbital and radial quantum numbers but differ in quark spin and total angular momenta have identical radial wavefunctions. Thus, for example, $\uD$ and $\uD^*$ have the same radial wavefunctions, as do $\chi_0$, $\chi_1$, $\chi_2$ and $h$. In that case the spatial matrix element $A_l$ is independent of the spin and total angular momenta, and the dependence on these quantum numbers only enters in the energy denominator and the coefficient $C_{SLJ}^{s_1s_2l}$.

In the limit in which the loop theorem of Barnes and Swanson \cite{Barnes:2007xu} is derived, all channels are characterised by a common reduced mass $\mu$ and binding energy $\epsilon$, and everything can be written in terms of common integrals,
\be
\Delta M^l	=\int dp \frac{p^2 |A_l(p)|^2}{\epsilon+{p^2}/{2\mu}}
,\quad
P^l		=\int dp \frac{p^2 |A_l(p)|^2}{\left(\epsilon+{p^2}/{2\mu}\right)^2}.
\label{meanintegrals}
\ee
Since the integrals are common to all channels the summations \rf{spinsums} involve only the recoupling factor, and from Table \ref{metable},
\be
\sum_{s_1s_2}C_{SLJ}^{s_1s_2l}=1,\label{spinunity}
\ee
which leads to the loop theorem.

To account for departures from the equal mass limit, consider a situation in which the bare masses of the quarkonia states are given by the perturbative formula \rf{mass}, while those in the loop are given by corresponding formula 
\be
m_{s_1,s_2}=m+\delta \evhh_{s_1,s_2}
\ee
and are characterised by centre-of-mass $m$ and splitting $\delta$. 

To find relations among the mass shifts, one can set up a power series expansion in which everything is expressed in terms of spin-averaged quantities, which correspond to setting all spin splittings to zero  ($\mu=m/2$ and $\epsilon=2m-M$). For a given channel the reduced mass and binding energy are related to their spin-averaged counterparts by
\bea
\frac{\mu_{s_1s_2}}{\mu}&=&1+\frac{\delta}{2m}\left(\evhh_{s_1}+\evhh_{s_2}\right)+O\left(\frac{\delta^2}{m^2}\right)\\
\epsilon_{SLJ}^{s_1s_2}&=&\epsilon +\delta\left(\evhh_{s_1}+\evhh_{s_2}\right)
-\Delta_{SLJ}.
\eea
Defining $X_{SLJ}^{s_1s_2}$ by
\be
\mu_{s_1s_2}\epsilon_{SLJ}^{s_1s_2}=\mu\epsilon(1+X_{SLJ}^{s_1s_2}),
\ee
one can express the mass shift as a power series 
\begin{multline}
\Delta M_{SLJ}^{s_1s_2l}= 
C_{SLJ}^{s_1s_2l}\frac{\mu_{s_1s_2}}{\mu}\frac{1}{\epsilon} 
\\
\times \sum_{n=0}^\infty(-X_{SLJ}^{s_1s_2})^n\int dp \frac{p^2 |A_l(p)|^2}{(1+{p^2}/{2\mu\epsilon})^{n+1}}.
\label{expansion}
\end{multline}
The series converges due to the smallness of $X_{SLJ}^{s_1s_2}$ and the monotonic decrease of the integrals with $n$. The results of BS \cite{Barnes:2007xu}, K \cite{Kalashnikova:2005ui} and LD \cite{Liu:2011yp}, with whose decay amplitudes the integrals above can be computed analytically, are very accurately reproduced keeping only the first three terms in the expansion. Keeping instead only the first two terms (which serves as a reasonable approximation for $\cc$ and an excellent approximation for $\bb$) the mass shift can be expressed in terms of the spin-averaged values $\Delta M^l$ and $P^l$ of equation \rf{meanintegrals},
\be
\Delta M_{SLJ}^{s_1s_2l}\approx C_{SLJ}^{s_1s_2l}\frac{\mu_{s_1s_2}}{\mu}\left(\Delta M^l-X_{SLJ}^{s_1s_2}\epsilon P^l\right).
\label{linearexpansion}
\ee
Ignoring terms suppressed by higher powers in the small parameter $\delta/m$, the mass shift is
\be
\Delta M_{SLJ}^{s_1s_2l}=C_{SLJ}^{s_1s_2l}\left(\Delta M^l+\Delta_{SLJ}P^l\right)
+\widetilde{C}_{SLJ}^{s_1s_2l}\delta Y^l
\ee
where
\bea
\widetilde{C}_{SLJ}^{s_1s_2l}&=&C_{SLJ}^{s_1s_2l}\left(\evhh_{s_1}+\evhh_{s_2}\right)\textrm{, and}\\
Y^l&=&\left(\frac{\Delta M^l}{2m}-\left(\frac{\epsilon}{2m}+1\right)P^l\right).
\eea
The total mass shift is the sum over those due to individual spin channels, as in equation \rf{spinsums}. The sum over the first term above is trivial on account of equation \rf{spinunity}. The sum over the second can be done using the coefficients in Table \ref{metable}, and the results are also shown there; remarkably  the dependence on $S$ and $J$  of the sum  is proportional to the matrix element of the spin-orbit operator,
\be
\sum_{s_1s_2}\widetilde{C}_{SLJ}^{s_1s_2l}=\xi_l\evso_{SLJ}
\ee
with $\xi_{\uS}=+1/2$  and $\xi_{\uD}=-1/4$. Thus the mass shift can be written
\be
 \Delta M_{SLJ}=\sum_l \left(\Delta M^l+\Delta_{SLJ}P^l
+ \delta\evso_{SLJ} \xi_l Y^l\right)
\label{final}
\ee
The correction to the hyperfine splitting due to loops follows immediately; everything cancels except a term proportional to $\Delta_s$,
\begin{multline}
-\frac{1}{9}\left(\Delta M_{\an 3P0}+3\Delta M_{\an 3P1}+5\Delta M_{\an 3P2}\right)+\Delta M_{\an 1\uP1}\\=-\Delta_s\sum_l P^l.
\end{multline}
Thus to this order, in the nonrelativistic limit ($\Delta_s=0$) the result of zero hyperfine splitting survives loop corrections. Away from this limit the hyperfine splitting, which is in any case inherently small, is actually reduced with respect to its quenched value. In most of the quoted examples in Table \ref{shiftstable}, there remains a small hyperfine splitting after loop corrections; this is due to quadratic corrections to the expansion \rf{linearexpansion}. The smallness of $X_{SLJ}^{s_1s_2}$ explains why the mechanism works even better for $\bb$ than $\cc$.

In equation \rf{final}, the first term is dominant and sets the scale of the mass shifts. The smallness of the latter two terms is a generic feature of any reasonable model, since $P^l$ is small, and $\Delta M^l,\epsilon << 2m$ so that $Y^l$ is also small. In the loop-induced spin-splitting between any two states,
\be
  \Delta M_{S'L'J'}- \Delta M_{SLJ},
\ee
the large term cancels, which underlines the observation that relative mass shifts are much smaller than overall mass shifts \cite{Ono:1983rd,Barnes:2007xu}.  

Thus there is a hierarchy of scales in the problem. While overall mass shifts due to loops can be large, the induced spin splittings are small, and the induced hyperfine splittings smaller still. The quark model prediction of small hyperfine splitting is thus robust against corrections due to unquenching. 

If one repeats the entire exercise for D-wave and higher $L$ mesons one finds that the same mechanism applies. Thus the prediction \cite{Burns:2010qq} for the mass of the $\an 1D2$ bottomonium in terms of the recently discovered \cite{delAmoSanchez:2010kz} $\an 3D{1,2,3}$ ought to be reliable.

The mechanism depends critically on the coefficients $C_{SLJ}^{s_1s_2l}$. These are common to all models in which quark spin and spatial degrees of freedom factorise, the coupling is driven by the creation of a spin-one pair, and the spin degrees of freedom are conserved. The observed small hyperfine splittings can thus be interpreted as support for this dynamical picture, which has already some support from lattice QCD calculations of strong decay \cite{Burns:2006wz}.

Table \ref{shiftstable} suggests two other general properties of loop-induced mass shifts. Firstly, the induced hyperfine splitting is always negative. If the physical hyperfine splitting is positive, as is favoured by the bulk of experimental and lattice data, then in the absence of some other effect the only possibility is that the potential model splitting $\Delta_s$ is positive. This may help to distinguish among different models, which disagree on the sign of $\Delta_s$ \cite{Godfrey:2002rp}. 

Secondly, the pattern of mass shifts is always the same,
\be
\Delta M_{\an 3P2}>\Delta M_{\an 1\uP1}>\Delta M_{\an 3P1}>\Delta M_{\an 3P0},
\ee
and implies the effect of unquenching is to bring the masses closer together with respect to their potential model values. (It turns out that this is not unique to P-wave levels.) The calculations of Eichten \etal \cite{Eichten:2004uh} exhibit a different pattern of splittings, due to mass shifts from mixing between different canonical configurations. Although in their case the induced hyperfine splitting remains small, this is not necessarily true of all models with configuration mixing. 

Note also that the calculations of Pennington and Wilson \cite{Pennington:2007xr} exhibit a large induced hyperfine splitting; their mass shifts are not protected by the mechanism outlined here because they do not sum over all spin combinations in the loop. The same applies to the simplified approach of Shmatikov \cite{Shmatikov:1998dy}.

The expression \rf{final} implies that the effect of unquenching can be absorbed into a renormalisation of $M$, $\Delta_s$, $\Delta_t$ and $\Delta_o$, and this may have interesting phenomenological consequences. For example, $\Delta_t$ and $\Delta_o$ are directly related to the relative contribution of vector and scalar parts of the interquark potential \cite{Voloshin:2007hh,Shmatikov:1998dy}. 

The factorisation of the angular momentum dependence of the mass shift by means of the expansion \rf{expansion} is similar in spirit to the approach of T\"ornqvist and Zenczykowski \cite{Tornqvist:1984fy}, although the expansion parameters differ. The advantage of the choice in this paper is that the first two terms in the expansion involve the average mass shift and continuum probability, both of which are physically meaningful quantities. The latter influences various hadron transition properties, such as radiative and pion decay widths. It may be possible, by means of the formalism presented here, to relate these properties the pattern of spin splittings, without making reference to any particular model.


\begin{acknowledgments}
The author thanks F. Close for useful discussions and C. Thomas for pointing out some lattice QCD references.
\end{acknowledgments}

\bibliography{tjb}

\end{document}

%% file: definitions.tex
\def\evhh{\ev{\tfrac{1}{2}\tfrac{1}{2}}}
\def\evso{\ev{\mathbf{L}\cdot\mathbf{S}} }
\def\evt{\ev{\mathbf T}}

\def\etal{$et$ $al$.}

\newcommand{\rf}[1]{(\ref{#1})}

\def\etab{\end{tabular}}  
  
\def\bit{\begin{itemize}}
\def\eit{\end{itemize}}

\def\bml{\begin{multline}}
\def\eml{\end{multline}}
\def\be{\begin{equation}} 
\def\ee{\end{equation}}  
\def\bea{\begin{eqnarray}}  
\def\eea{\end{eqnarray}} 
\def\bes{\begin{equation*}} 
\def\ees{\end{equation*}}  
\def\beas{\begin{eqnarray*}}  
\def\eeas{\end{eqnarray*}} 
\def\bmu{\begin{multline}}
\def\emu{\end{multline}}
\def\bal{\begin{array}{l}} 	
\def\eal{\end{array}}

\def\MeV{\textrm{ MeV}}

\def\letterS{S}
\def\letterP{P}
\def\letterD{D}
\def\letterF{F}
\def\greek#1{\def\letter{#1}
\ifx\letter\letterS\Sigma
\else
	\ifx\letter\letterP\Pi
	\else
		\ifx\letter\letterD\Delta
		\else
			\ifx\letter\letterF\Phi
			\else XXX
			\fi
		\fi
	\fi
\fi
}

\def\cc{c\overline c}

\def\cc{c\overline c}

\def\bb{b\overline b}

\def\bb{b\overline b}

\newcommand{\ket}[1]{|           {#1}           \rangle}
\newcommand{\bra}[1]{\langle           {#1}           |}

\newcommand{\ev}[1]{\langle #1 \rangle}

\renewcommand{\u}[1]{\textrm{#1}}
\def\cn#1#2#3{^{#1}\u {#2}_{#3}}     
\def\an{\cn} 
\newcommand{\uS}{\textrm{S}}
\newcommand{\uP}{\textrm{P}}
\newcommand{\uD}{\textrm{D}}

\def\epem{e^+e^-}

\def\xQN#1,{\def\Element{#1}%
\ifx\Element\endpiece
\else 
	\if\Element/\def\ApplyOrLiteral{1} 
	\else
		\if\ApplyOrLiteral0\ElementRule\Element 
		\else\Element\def\ApplyOrLiteral{0}
		\fi
	\fi
	\expandafter\xQN
\fi}

\def\xNQ#1,{\def\ElementRule{#1}%
\ifx\ElementRule\endpiece
\else 
	\if\ElementRule/\def\ApplyOrLiteral{1} 
	\else
		\if\ApplyOrLiteral0\ElementRule\Element 
		\else\ElementRule\def\ApplyOrLiteral{0}
		\fi
	\fi
	\expandafter\xNQ
\fi}

\def\k|#1>{\ket{#1}} 
\def\b<#1|{\bra{#1}} 

\def\bk<#1|#2>{\langle #1 | #2 \rangle}
\def\Bk#1<#2|#3>{\def\ElementRule{#1}\def\ApplyOrLiteral{0}
\xQN/,\langle,#2,xxx,\vert #3 \rangle}
\def\bK#1<#2|#3>{Undefined}

\def\BK#1<#2|#3>{\def\ElementRule{#1}\def\ApplyOrLiteral{0}
\xQN/,\langle,#2,/,|,#3,/,\rangle,xxx,}
\def\KB<#1|#2>#3{\def\Element{#3}\def\ApplyOrLiteral{0}
\xNQ/,\langle,#1,/,|,#2,/,\rangle,xxx,}

\def\me<#1|#2|#3>{\langle #1\vert #2\vert #3\rangle}
\def\ME#1<#2|#3|#4>{\def\ElementRule{#1}\def\ApplyOrLiteral{0}
\xQN/,\langle,#2,/,\vert,/,#3,/,\vert,#4,/,\rangle,xxx,}
\def\EM<#1|#2|#3>#4{\def\Element{#4}\def\ApplyOrLiteral{0}
\xNQ/,\langle,#1,/,\vert,/,#2,/,\vert,#3,/,\rangle,xxx,}

\def\rme<#1|#2|#3>{\langle #1\Vert #2\Vert #3\rangle}
\def\RME#1<#2|#3|#4>{\def\ElementRule{#1}\def\ApplyOrLiteral{0}
\xQN/,\langle,#2,/,\Vert,/,#3,/,\Vert,#4,/,\rangle,xxx,}
\def\EMR<#1|#2|#3>#4{\def\Element{#4}\def\ApplyOrLiteral{0}
\xNQ/,\langle,#1,/,\Vert,/,#2,/,\Vert,#3,/,\rangle,xxx,}

\def\endpiece{xxx}

